# A Multivariate Density Forecast Approach for Online Power System Security Assessment

Zichao Meng, Ye Guo, *Senior Member, IEEE*, Wenjun Tang, *Member, IEEE*, Hongbin Sun, *Fellow, IEEE*, and Wenqi Huang

*Abstract*—A multivariate density forecast model based on deep learning is designed in this paper to forecast the joint cumulative distribution functions (JCDFs) of multiple security margins in power systems. Differing from existing multivariate density forecast models, the proposed method requires no a priori hypotheses on the distribution of forecasting targets. In addition, based on the universal approximation capability of neural networks, the value domain of the proposed approach has been proven to include all continuous JCDFs. The forecasted JCDF is further employed to calculate the deterministic security assessment index evaluating the security level of future power system operations. Numerical tests verify the superiority of the proposed method over current multivariate density forecast models. The deterministic security assessment index is demonstrated to be more informative for operators than security margins as well.

## I. INTRODUCTION

### A. Background

With the increasing integration of renewable energies and power electronic devices, power system operations have been more uncertain and volatile than ever before [1]. This brings significant challenges to the secure operation of power systems. Therefore, a more reliable, more adaptive, smarter, and faster approach for power system security assessment is of crucial importance.

Unfortunately, with more and more new types of power generation and transmission devices, obtaining an accurate model for the real-time security assessment has also become more and more challenging. As a result, the performance of traditional model-driven approaches gets deteriorated. Researchers thus have been focusing on the data-driven method as an alternative.

### B. Literature Review

Data-driven methods can extract operation security knowledge from data in a model-free manner and respond very quickly to different operation scenarios [2]. A crucial criteria used in security assessment is the security margin. Data-driven methods with this topic can be categorized into two types: point estimation and density estimation methods.

In point estimation methods, we predict the expectation of security margins. Various machine learning algorithms have been applied to the evaluation of security margins or security regions in the power system, including ensemble learning [3], group loss learning [4], multitask learning [5], and hybrid deep learning [6]. Benefiting from the high computational performance of distributed computing platforms [7], the computational efficiencies of these methods can be significantly improved.

In density estimation methods, we predict the probability density function (PDF) or the cumulative distribution function (CDF) of security margins. Several kinds of indicators related to the security margin in power systems have been investigated in the literature, such as the load margin [8], available transfer capability (ATC) [9], and total transfer capability (TTC) [10]-[11]. Specifically, authors of [8] proposed a Gaussian-process-emulator-based approach to assess probabilistic distributions of load margins. Paper [9] developed a parametric bootstrap technique to estimate the distribution information of ATC. In [10], a polynomial chaos expansion algorithm was proposed for the probabilistic TTC assessment, which was further improved by authors of [11] with sparse techniques to make the approach more efficient.

Similar to density estimation, density forecast approaches predict the future PDF or CDF of the forecasting target [12]. They are meaningful to the early warning of power system security risks, yet have been rarely utilized in the security assessment so far. Existing density forecast models focus on the univariate situation [13]-[15], among which distribution approximation network-network forecast network (DAN-NFN) [13] is one of the most effective models for the forecast of wind power and electricity prices. In DAN-NFN, no a priori hypotheses are introduced on the forecasted distribution and well-estimated results of the forecasted distribution are obtained based on deep neural networks (NNs). However, univariate density forecast models, such as DAN-NFN, fail to demonstrate the dependency among different variables. In real power systems, operators should simultaneously monitor security margins of multiple transmission interfaces, multivariate density forecast models are thus indispensable for online security assessment with uncertainties. The design of

This work was supported by the National Key R&D Program of China (2020YFB0906000, 2020YFB0906005). *(corresponding author: Ye Guo.)*

Z. Meng (mzc20@mails.tsinghua.edu.cn), Y. Guo (guo-ye@sz.tsinghua.edu.cn), and W. Tang are with the Smart Grid and Renewable Energy Lab, Tsinghua-Berkeley Shenzhen Institute (TBSI), Tsinghua University, Shenzhen 518055, China.

H. Sun (shb@tsinghua.edu.cn) is with the Department of Electrical Engineering, State Key Laboratory of Power Systems, Tsinghua University, Beijing 100084, China, and also with the Smart Grid and Renewable Energy Lab, Tsinghua-Berkeley Shenzhen Institute (TBSI), Tsinghua University, Shenzhen 518055, China.

W. Huang is with the Digital Grid Research Institute, China Southern Power Grid, Guangzhou 510670, China.

multivariate density forecast models is a tough task with few relevant investigations. Current multivariate density forecast models are mainly based on multivariate kernel density estimation (MKDE) [16] or higher dimensional copula [17]. Compared with DAN-NFN, extra limitations on the forecasted joint distribution are introduced in these models. For example, for MKDE-based models, the joint PDFs (JPDFs) of their forecasted distributions have to be the sum of a finite number of kernel functions as the number of historical samples is always finite. For copula-based ones, their forecasted distributions have to conform to copula families with hyper-parameters. Due to the finiteness of available datasets (in MKDE-based models) and a priori hypotheses of copula families (in copula-based models), the forecasted joint distributions may not cover all the possible real joint distributions. Thus, potential optimal distributions may be missed, and these methods' abilities to approximate the real joint distribution of forecasting targets are limited.

*C. Contributions*

The main contribution of this work is twofold. First, a novel data-driven multivariate density forecast model is proposed to estimate the joint CDFs (JCDFs) of future security margins. The proposed model is the multivariate form of DAN-NFN and consists of two deep NNs: a joint distribution approximation network (JDAN) and a network forecast network (NFN). The advantages of JDAN-NFN, compared with previous works in [13]-[17], include: 1) there are no a priori hypotheses on the forecast of joint distributions, i.e., the proposed approach is distribution-free and 2) JDAN-NFN can be regarded as an ideal multivariate density forecast model when its capacity is large enough, which brings the potential for finding more satisfied forecasted results than existing multivariate density forecast models.

Second, based on the forecasted JCDFs from JDAN-NFN, a deterministic security assessment index $\Omega$ is developed to indicate the future security of power system operations, which is more informative than the security margin.

The rest of this paper is organized as follows. Section II presents preliminaries about density forecast, DAN-NFN, and the joint distribution of security margins. Section III introduces the model framework. Section IV details the model building and training. Section V presents numerical simulations results, and conclusions are drawn in Section VI.

## II. PRELIMINARY

*A. Multivariate Density and Ideal Density Forecast Models*

Multivariate density forecast aims to obtain the joint distribution of multiple future random variables (forecasting targets) based on information available. Denoting the available information set up to time $t$ as $X_t \in E$ ($E$ is the value domain of $X_t$), multivariate density forecast with lead step $\tau$ aims to estimate the real joint distribution of the forecasting target $y_{t+\tau}$ at time spot $t+\tau$ (denoted as $\phi_{t+\tau}|_{X_t}$), i.e. $y_{t+\tau} \sim \phi_{t+\tau}|_{X_t}$. The distribution $\phi_{t+\tau}|_{X_t}$ can take the form of JPDF $f(\cdot|X_t)$ or JCDF $F(\cdot|X_t)$, and $\phi_{t+\tau}|_{X_t}$ is considered as JCDF if there is no additional instruction in this paper.

A multivariate density forecast model can be regarded as a function $\hat{\phi}_{t+\tau}|_{X_t} = \Psi(X_t; \theta)$, where $\Psi(\cdot; \theta)$ is the density forecast model and $\theta$ includes parameters of $\Psi$. $\hat{\phi}_{t+\tau}|_{X_t}$ is the forecasted joint distribution function, which is an estimate of $\phi_{t+\tau}|_{X_t}$. Different $\hat{\phi}_{t+\tau}|_{X_t}$ can be determined by changing $\theta$. The training of $\Psi(\cdot; \theta)$ is to find the best $\theta$, so that the forecasted $\hat{\phi}_{t+\tau}|_{X_t}$ is as close to $\phi_{t+\tau}|_{X_t}$ as possible.

We next introduce the notion of *ideal density forecast model* [13]. Given any $X_t$, the value domain of $\Psi$ for different $\theta$ is denoted by $D_{X_t}$, and we call it the forecast domain of $\Psi$ at $X_t$. Similarly, the feasible domain of all the possible $\phi_{t+\tau}|_{X_t}$ is defined as $\Xi$. $\Psi$ is called an *ideal density forecast model* if
$$\forall X_t \in E, \ D_{X_t} = \Xi. \quad (1)$$

For non-ideal density forecast models (e.g. MKDE-based and copula-based models), their forecast domains may not be equal to the feasible domain. Specifically, in MKDE-based models, the forecasted JPDF is a sum of kernel functions. Since the number of kernel functions is equal to that of historical samples and is always finite, the forecasted joint distributions may not cover all the possible joint distributions in the feasible domain. Thus, the real joint distribution may not be included in the forecast domain. In copula-based models, the forecasted joint distribution should obey the copula family with hyper-parameters. Due to these limitations, these models may not be an *ideal density forecast model* with the given copula family, thus may miss the real joint distribution as well. It can be inferred that the forecast domain of the *ideal density forecast model* always includes the real joint distribution. Thus, the ideal model possesses the attractive property of having the potential for finding the real joint distribution.

*B. Basic Structure of DAN-NFN*

Our method is developed based on DAN-NFN, so the basic structure of DAN-NFN is firstly introduced. There are two deep NNs in DAN-NFN, namely, DAN and NFN. DAN is a single-in-single-out (SISO) positive-weighted artificial neural network (ANN) and is built for approximating the real CDFs of the forecasting target. For NFN, it takes available information set as input and outputs all parameters of DAN. Based on the nonlinear approximating capability of NNs [18] and the viewpoint that a SISO positive-weighted ANN can approximate any univariate continuous monotone non-decreasing function [19], DAN-NFN can be regarded as an *ideal density forecast model* if its capacity is built large enough, which makes it superior over other existing univariate density forecast models. However, the application of deep learning approaches to multivariate density forecast is still an open question.

*C. Joint Distribution of Multiple Security Margins*

Different areas of a large power system are usually connected by tie-lines, and a group of which is referred to as a "transmission flowgate". The sum of power flow through a flowgate should not exceed its capacity limit, which is usually referred to as TTC. The security margin of a flowgate $i$ is defined as
$$SM^i = 1 - P^i/P^i_{ttc}, \quad (2)$$
where $P^i$ denotes the transmission power of flowgate $i$, $P^i_{ttc}$ denotes the corresponding TTC. Owing to the uncertainties in the power system, the joint distribution of ***SM*** in a power system is a crucial basis for its security assessment.

Information set of spatial-temporal state variables at *t* and before is collected and normalized as $\boldsymbol{X}_t$. More specifically, $\boldsymbol{X}_t$ is an information matrix whose columns represent different features (data measured in the power system), and rows represent different time spots. Given lead step $\tau$ and the forecasting target $\boldsymbol{SM}_{t+\tau}$, the multivariate density forecast function is designed to estimate the joint distribution $\boldsymbol{\phi}_{t+\tau}|_{\boldsymbol{X}_t}$ of $\boldsymbol{SM}_{t+\tau}$. Assuming that there are *N* flowgates in the power system, the feasible domain $\Xi$ of $\boldsymbol{\phi}_{t+\tau}|_{\boldsymbol{X}_t}$ with JPDF $f(\cdot|\boldsymbol{X}_t)$ can be denoted as

$$\Xi = \left\{ \boldsymbol{\phi}_{t+\tau}|_{\boldsymbol{X}_t} \text{ with } \middle| \begin{array}{l} f(\cdot|\boldsymbol{X}_t) \text{ is nonnegtive and bounded,} \\ \text{JPDF} f(\cdot|\boldsymbol{X}_t) \middle| \int_{-\infty}^{+\infty}\cdots\int_{-\infty}^{+\infty} f(\boldsymbol{SM}_{t+\tau}|\boldsymbol{X}_t)d SM_{t+\tau}^1 \cdots d SM_{t+\tau}^N = 1 \end{array} \right\}.$$
(3)

Subsequently, $\boldsymbol{\phi}_{t+\tau}|_{\boldsymbol{X}_t}$ with JCDF $F(\cdot|\boldsymbol{X}_t)$ can be written as

$$F(\boldsymbol{SM}_{t+\tau}|\boldsymbol{X}_t) = \int_{-\infty}^{SM_{t+\tau}^1}\cdots\int_{-\infty}^{SM_{t+\tau}^N} f(\boldsymbol{SM}_{t+\tau}|\boldsymbol{X}_t) dSM_{t+\tau}^1 \cdots dSM_{t+\tau}^N.$$
(4)

$f(\cdot|\boldsymbol{X}_t)$ being bounded leads to $F(\cdot|\boldsymbol{X}_t)$ being continuous. The limit of $F(\boldsymbol{SM}_{t+\tau}|\boldsymbol{X}_t)$ is equal to 1 as $\boldsymbol{SM}_{t+\tau}$ approaches $+\infty$ and is equal to 0 as any variable in $\boldsymbol{SM}_{t+\tau}$ approaches $-\infty$. Thus, from the perspective of JCDF, $\Xi$ can be rewritten as

$$\Xi = \left\{ \boldsymbol{\phi}_{t+\tau}|_{\boldsymbol{X}_t} \text{ with } \middle| \begin{array}{l} F(\cdot|\boldsymbol{X}_t) \text{ is continous,} \\ F^{(N)}(\boldsymbol{SM}_{t+\tau}|\boldsymbol{X}_t) \geq 0, \\ \lim_{\boldsymbol{SM}_{t+\tau}\to+\infty} F(\boldsymbol{SM}_{t+\tau}|\boldsymbol{X}_t) = 1, \\ \lim_{SM_{t+\tau}^i\to-\infty} F(\boldsymbol{SM}_{t+\tau}|\boldsymbol{X}_t) = 0, \ \forall i \in [1,N] \end{array} \right\},$$
(5)

where $F^{(N)}(\boldsymbol{SM}_{t+\tau}|\boldsymbol{X}_t) = \dfrac{\partial^N (F(\boldsymbol{SM}_{t+\tau};\boldsymbol{W}^+,\boldsymbol{B}))}{\partial SM_{t+\tau}^1 \cdots \partial SM_{t+\tau}^N}$.

## III. MODEL FRAMEWORK

The framework of power system security assessment is demonstrated in Fig. 1. First, JDAN-NFN is designed to forecast the joint distribution of multiple security margins. Such forecasts will further be translated to estimates of the deterministic security assessment index.

### A. JDAN-NFN

#### 1) Structure of JDAN-NFN

JDAN-NFN multivariate density forecast model is the core part in the framework of power system security assessment. It estimates the joint distribution of $\boldsymbol{SM}_{t+\tau}$ (namely, $\boldsymbol{\phi}_{t+\tau}|_{\boldsymbol{X}_t}$), which is represented by its JCDF $F(\cdot|\boldsymbol{X}_t)$. Before describing the structure of JDAN-NFN, one approximation theorem for ANN is introduced below.

*Theorem 1* [19]: Let $\mathcal{X}$ be an input space represented by *M* attributes, for any continuous monotone non-decreasing function $f : \aleph \to \mathbb{R}$, where $\aleph$ is a compact subset of $\mathbb{R}^M$, there exists a feedforward neural network with at most *M* hidden layers, positive weights, and output $\Gamma$ such that $|f(x) - \Gamma_x| < \varepsilon$, for any $x \in \mathcal{X}$ and $\varepsilon > 0$.

As the JCDF of security margins is a multivariate monotone function, a multiple-in-single-out (MISO) positive-weighted ANN is adopted to approximate it. From this point of view, we still adopt a similar idea as DAN-NFN: NFN is used to generate the parameters of JDAN, which guarantees JDAN's weights are positive, and JDAN is used to approximate the JCDF.

To be specific, NFN outputs Tensors $\boldsymbol{W}^+$ and $\boldsymbol{B}$ with different activation function. Tensor $\boldsymbol{W}^+$ comes from SoftPlus function $SP(x) = \ln(1+e^x)$, thus $\boldsymbol{W}^+ \in \mathbb{R}^+$. Tensor $\boldsymbol{B}$ comes from linear function, thus $\boldsymbol{B} \in \mathbb{R}$. In short, NFN maps input $\boldsymbol{X}_t$ to $[\boldsymbol{W}^+,\boldsymbol{B}]$. JDAN is a MISO ANN with multiple hidden layers. Picking $[\boldsymbol{W}^+,\boldsymbol{B}]$ as parameters of JDAN, the input-output mapping function of JDAN can be regarded as a deterministic function $\Phi_J(\cdot;\boldsymbol{W}^+,\boldsymbol{B})$.

#### 2) Estimation of JCDF

From subsection II-C, we know that a reasonable estimation of JCDF $F(\cdot|\boldsymbol{X}_t)$ [denoted as $\hat{F}(\cdot|\boldsymbol{X}_t)$] should satisfy the following conditions in (5):

(i) $\hat{F}(\cdot|\boldsymbol{X}_t)$ is continuous,

(ii) $\hat{F}^{(N)}(\boldsymbol{SM}_{t+\tau}|\boldsymbol{X}_t) \geq 0$,

(iii) $\lim_{\boldsymbol{SM}_{t+\tau}\to+\infty} \hat{F}(\boldsymbol{SM}_{t+\tau}|\boldsymbol{X}_t) = 1$,

(iv) $\lim_{SM_{t+\tau}^i\to-\infty} \hat{F}(\boldsymbol{SM}_{t+\tau}|\boldsymbol{X}_t) = 0, \ \forall i \in [1,N]$.

It can be proven that a MISO positive-weighted ANN is a multivariate monotone non-decreasing function. However, one cannot extend this property to the higher-order-derivative form that meets condition (ii), which is demonstrated in Appendix A. Therefore, a normal MISO positive-weighted ANN cannot be used to represent the JCDF. To address this problem, we propose a novel structure of positive-weighted ANN, i.e. JDAN, which consists of four parts as shown in Fig. 2:

1) Parallel units

Each variable in $\boldsymbol{SM}_{t+\tau}$ is sent into a SISO positive-weighted ANN (can be regarded as a parallel unit as well), respectively. Sigmoid function $\sigma(x) = 1/(1+e^{-x})$ (denote as $\sigma$) is chosen as the activation of hidden layers and linear function is chosen as the activation of output layer in each parallel unit, i.e.

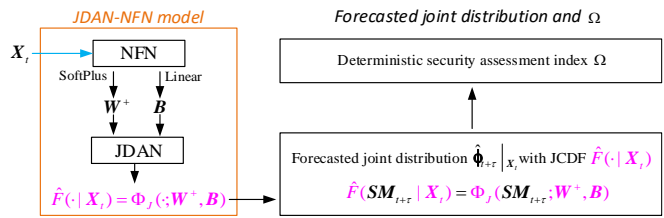

Fig. 1. Framework of power system security assessment (purple notations represent not variables but functions).

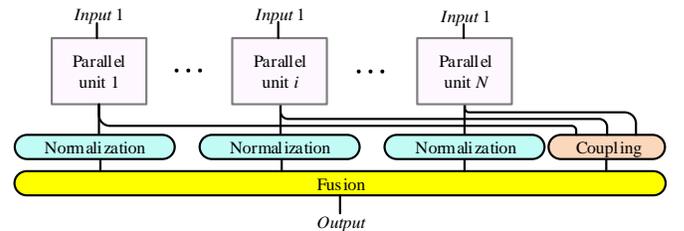

Fig. 2. Framework of JDAN.

$$\Phi^i(SM_{t+\tau}^i; W^{i+}, B^i) = W_{K+1}^{i+} \cdots \sigma[W_2^{i+} \cdot \sigma(SM_{t+\tau}^i \cdot W_1^{i+} + B_1^i) + B_2^i] \cdots + B_{K+1}^i, \quad (6)$$

where $SM_{t+\tau}^i$ is the $i$th variable in $SM_{t+\tau}$. Tensors $W^{i+}$ and $B^i$ represent all the weights and biases in the $i$th parallel unit, respectively, which are determined by the outputs of NFN. $K$ is the number of hidden layers in the $i$th parallel unit. Tensors $W_k^{i+} \in W^{i+}$ and $B_k^i \in B^i$, $k \in [1, K+1]$, are weights and biases in the $k$th layer of the $i$th parallel unit, respectively. $\Phi^i(\cdot; W^{i+}, B^i)$ is the input-output mapping function of the $i$th parallel unit, and the number of parallel units is equal to the dimension of $SM_{t+\tau}$. One can prove that (6) is a monotonically increasing function in the same way [following (31)-(37)] demonstrated in Appendix A.

2) Normalization Layers

The output of each parallel unit will be normalized by a normalization layer, denoted as

$$\overline{\Phi}^i = \frac{\Phi^i(SM_{t+\tau}^i; W^{i+}, B^i) - \lim_{L_i \to -\infty} \Phi^i(L_i; W^{i+}, B^i)}{\lim_{U_i \to +\infty} \Phi^i(U_i; W^{i+}, B^i) - \lim_{L_i \to -\infty} \Phi^i(L_i; W^{i+}, B^i)}, \quad (7)$$

where $U_i$ and $L_i$ are the upper and lower bounds of $SM_{t+\tau}^i$, respectively. One can verify that $\overline{\Phi}^i$ is also monotonously non-decreasing, and $\lim_{SM_{t+\tau}^i \to +\infty} \overline{\Phi}^i = 1$, $\lim_{SM_{t+\tau}^i \to -\infty} \overline{\Phi}^i = 0$.

3) Coupling Layer

Coupling layer is used to construct the coupling relationship among input variables of JDAN based on the output of the parallel unit. In this paper, the coupling layer is represented as

$$\widehat{\Phi} = \frac{\sum_{i=1}^{N}[\Phi^i(SM_{t+\tau}^i; W^{i+}, B^i) - \lim_{L_i \to -\infty} \Phi^i(L_i; W^{i+}, B^i)]}{\sum_{i=1}^{N}[\lim_{U_i \to +\infty} \Phi^i(U_i; W^{i+}, B^i) - \lim_{L_i \to -\infty} \Phi^i(L_i; W^{i+}, B^i)]}, \quad (8)$$

We can easily verify that $\widehat{\Phi}$ is multivariate non-decreasing as well, and $\lim_{SM_{t+\tau} \to +\infty} \widehat{\Phi} = 1$.

4) Fusion Layer

Fusion layer receives the results of all normalization layers and the coupling layer. The input-output mapping function of JDAN is finally constructed as

$$\Phi_J(SM_{t+\tau}; W^+, B) = \widehat{\Phi} \cdot \prod_{i=1}^{N} \overline{\Phi}^i, \quad (9)$$

Based on the monotonicity of (6), (7), and (8), one can verify that

$$\Phi_J^{(N)}(SM_{t+\tau}; W^+, B) \geq 0. \quad (10)$$

From the product law of limitations, we have the following limits as

$$\lim_{SM_{t+\tau} \to +\infty} \Phi_J(SM_{t+\tau}; W^+, B) = \lim_{SM_{t+\tau} \to +\infty} \widehat{\Phi} \cdot (\prod_{i=1}^{N} \lim_{SM_{t+\tau}^i \to +\infty} \overline{\Phi}^i) = 1, \quad (11)$$

$$\lim_{SM_{t+\tau}^i \to -\infty} \Phi_J(SM_{t+\tau}; W^+, B) = 0, \quad \forall i \in [1, N]. \quad (12)$$

Note that, $\Phi_J(SM_{t+\tau}; W^+, B)$ is continuous as the activation functions of JDAN are sigmoid and linear, thus meets condition (i). According to (10)-(12), conditions (ii)-(iv) are also satisfied. Eventually, the forecasted joint distribution $\hat{\phi}_{t+\tau}|_{X_t}$ with JCDF $\hat{F}(\cdot | X_t)$ is formulated as

$$\hat{F}(SM_{t+\tau} | X_t) = \Phi_J(SM_{t+\tau}; W^+, B). \quad (13)$$

The forecasted JPDF $\hat{f}(\cdot | X_t)$ can be derived from (13) as

$$\hat{f}(SM_{t+\tau} | X_t) = \Phi_J^{(N)}(SM_{t+\tau}; W^+, B). \quad (14)$$

In summary, to estimate JCDFs, a special MISO positive-weighted ANN (namely JDAN) is established, which includes parallel unit, normalization layer, coupling layer, and fusion layer. It should be noticed that the coupling layer may also be formulated in other ways, and when designing it, we should pay attention to not only conditions (i)-(iv), but also the issue of variable dependency. Otherwise, the model trained may not be able to demonstrate the dependency among variables, and an example for variable independency is illustrated in Appendix B. The detailed information of the parallel unit of JDAN, as well as NFN, will be introduced in section IV, and we have proven that JDAN-NFN can be regarded as an ideal multivariate density forecast model if its capacity is built large enough (see Appendix C).

B. A Deterministic Security Assessment Index $\Omega$

Now, we can estimate the joint distribution function of $SM_{t+\tau}$ through JDAN-NFN. In a very general way, $SM_{t+\tau}$ should be limited in a certain range setting by operators to maintain the secure operation of the power system, i.e.

$$P^i / P_{ttc}^i \leq \gamma_i, \quad \forall i \in [1, N], \quad (15)$$

where $\gamma_i$ is the security threshold of flowgate $i$, which can be determined by operators. According to (2), rewrite (15) in the form of security margins as

$$SM_{t+\tau}^i \geq 1 - \gamma_i, \quad \forall i \in [1, N]. \quad (16)$$

Then, we define the deterministic security assessment index $\Omega$ as

$$\Omega = \int_{1-\gamma_1}^{\infty} \cdots \int_{1-\gamma_N}^{\infty} \hat{f}(SM_{t+\tau} | X_t) dSM_{t+\tau}^1 \cdots dSM_{t+\tau}^N. \quad (17)$$

Take the univariate distribution for illustration, $\Omega$ is the integral of PDF with respect to $SM_{t+\tau}^i$ on the interval determined by (16) (the green area shown in Fig. 3). Considering a simple example of two variables [$(SM_{t+\tau}^1, SM_{t+\tau}^2)$] for (17), the corresponding deterministic security assessment index $\Omega_2$ can be rewritten based on the JCDF approximated by JDAN-NFN as

$$\Omega_2 = \hat{F}((+\infty, +\infty) | X_t) - \hat{F}((1-\gamma_1, +\infty) | X_t) - \hat{F}((+\infty, 1-\gamma_2) | X_t) + \hat{F}((1-\gamma_1, 1-\gamma_2) | X_t). \quad (18)$$

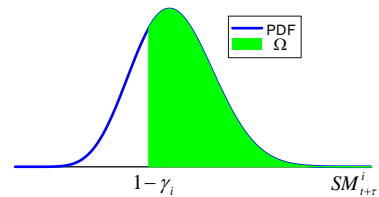

Fig. 3. $\Omega$ of univariate distribution.

Extend (18) to the form of $N$ variables, then we have

$$\Omega = \sum_{k=1}^{2^N} (-1)^{\lambda_k} F(\boldsymbol{\gamma}^k \mid \boldsymbol{X}_t), \quad \boldsymbol{\gamma}^k \in \boldsymbol{\gamma}, \quad (19)$$

where $\boldsymbol{\gamma}$ is the set of all possible combinations of lower and upper bounds of the $N$ variables as in (17). For each element $\boldsymbol{\gamma}^k$ in $\boldsymbol{\gamma}$, the number of entries associated with lower bounds of security margins is denoted by $\lambda_k$. From (18) or (19), we know that $\Omega \in [0, 1]$, and the smaller the $\Omega$, the greater the risk in power systems.

## IV. MODEL BUILDING AND TRAINING

In this section, more details of the JDAN-NFN are elaborated.

### A. Building NFN and the Parallel Unit of JDAN

The design of NFN and JDAN should make their capacities as large as possible, so that the JDAN-NFN will be an ideal multivariate density forecast model. To that end, deeper structures are adopted since they are more effective in increasing the capacities of NNs [18]. However, one challenge in training deep NNs is that parameters of deep layers may be updated very slowly. Thus, residual structure [20], which is realized through residual connections (identical mappings between nonadjacent layers), is implemented in both NFN and JDAN. Since the inputs of NFN are time series $\boldsymbol{X}_t$, long short-term memory (LSTM) network [21] is used in this paper for the modeling of temporal relations. Specifically, the LSTMs adopted here share the same deep residual structure as the one demonstrated in [13]. Given a lag interval $\delta$, the input matrix to LSTM can be written as $\boldsymbol{X}_t = [\boldsymbol{x}_{t-\delta+1}, \cdots, \boldsymbol{x}_t]^T$, then from time spot $\kappa = t - \delta + 1$ to $\kappa = t$ the recurrent calculation steps in LSTM are as follows:

$$\boldsymbol{i}_\kappa = \sigma(\boldsymbol{w}_{ix} \cdot \boldsymbol{x}_\kappa + \boldsymbol{w}_{ih} \cdot \boldsymbol{h}_{\kappa-1} + \boldsymbol{w}_{ic} \cdot \boldsymbol{c}_{\kappa-1} + \boldsymbol{b}_i), \quad (20)$$

$$\boldsymbol{f}_\kappa = \sigma(\boldsymbol{w}_{fx} \cdot \boldsymbol{x}_\kappa + \boldsymbol{w}_{fh} \cdot \boldsymbol{h}_{\kappa-1} + \boldsymbol{w}_{fc} \cdot \boldsymbol{c}_{\kappa-1} + \boldsymbol{b}_f), \quad (21)$$

$$\boldsymbol{c}_\kappa = \boldsymbol{f}_\kappa \odot \boldsymbol{c}_{\kappa-1} + \boldsymbol{i}_\kappa \odot \tanh(\boldsymbol{w}_{cx} \cdot \boldsymbol{x}_\kappa + \boldsymbol{w}_{ch} \cdot \boldsymbol{h}_{\kappa-1} + \boldsymbol{b}_c), \quad (22)$$

$$\boldsymbol{o}_\kappa = \sigma(\boldsymbol{w}_{ox} \cdot \boldsymbol{x}_\kappa + \boldsymbol{w}_{oh} \cdot \boldsymbol{h}_{\kappa-1} + \boldsymbol{w}_{oc} \cdot \boldsymbol{c}_{\kappa-1} + \boldsymbol{b}_o), \quad (23)$$

$$\boldsymbol{h}_\kappa = \boldsymbol{o}_\kappa \odot \tanh(\boldsymbol{c}_\kappa), \quad (24)$$

where $\boldsymbol{i}_\kappa$, $\boldsymbol{f}_\kappa$ and $\boldsymbol{o}_\kappa$ represent input gate, forget gate and output gate, respectively. Vectors $\boldsymbol{h}_\kappa$ and $\boldsymbol{c}_\kappa$ represent cell state and hidden state, respectively. The sizes (number of elements) of $\boldsymbol{h}_\kappa$ and $\boldsymbol{c}_\kappa$ are the same. Tensors $\boldsymbol{w}$ with different subscripts are weights, and $\boldsymbol{b}$ with different subscripts are biases. $\odot$ represents entrywise product. We define that the recurrent calculation steps (20)-(24) as one LSTM layer, and the size of $\boldsymbol{h}_\kappa$ or $\boldsymbol{c}_\kappa$ as its width.

As shown in Fig. 4(a), NFN is built by many LSTM layers with residual connections and two fully connected (FC) layers. Batch normalization (BN) [22] is also implemented to accelerate the training of NFN. For JDAN, as illustrated in Fig. 4(b), parallel units are constructed by many fully connected (FC) layers with residual connections. Multiple parallel units, normalization layers, and one fusion layer are placed in series. One coupling layer connects all parallel units and the fusion layer.

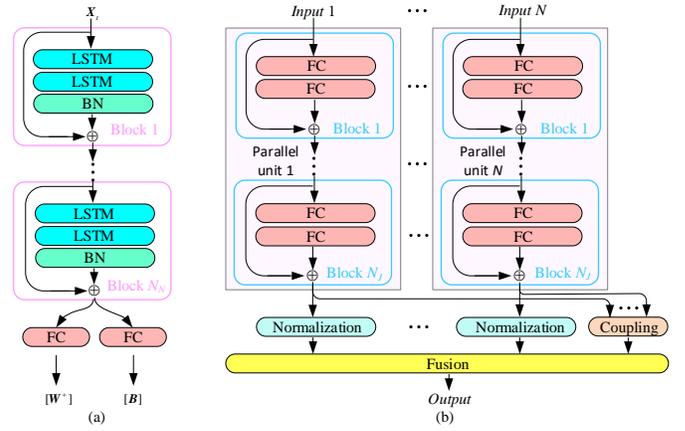

Fig. 4. Detail structures of NFN and JDAN. (a) NFN. (b) JDAN.

Fig. 4 shows that the main parts of NFN and the parallel units of JDAN are both composed of similar components, which contain several layers (LSTM or FC) and residual connections. Intuitively, such components may also be named as blocks. $N_N$ and $N_J$ are the numbers of blocks in NFN and JDAN, respectively. As we focus on the depth of NNs, the layers in NFN or JDAN are built with the same widths for simplicity, i.e., the widths of LSTM layers and FC layers are $W_N$ in NFN and the widths of FC layers are $W_J$ in JDAN.

### B. Training JDAN-NFN

*1) Objective Function*

According to (13)-(14), JDAN-NFN can estimate the JCDF $\hat{F}(\cdot \mid \boldsymbol{X}_t)$ of security margins and the corresponding JPDF $\hat{f}(\cdot \mid \boldsymbol{X}_t)$, then the objective function for maximum likelihood estimation (MLE) is constructed based on the JPDF. Denoting the observation (real measurement) of $\boldsymbol{SM}_{t+\tau}$ as $\boldsymbol{SM}_{t+\tau}^*$, the objective function $L(\boldsymbol{\theta}_N)$ used in the training process is formulated as

$$L(\boldsymbol{\theta}_N) = \ln[\hat{f}(\boldsymbol{SM}_{t+\tau}^* \mid \boldsymbol{X}_t)] = \ln[\Phi_J^{(N)}(\boldsymbol{SM}_{t+\tau}^*; \boldsymbol{W}^+, \boldsymbol{B})], \quad (25)$$

where $\boldsymbol{\theta}_N$ represents parameters of NFN.

*2) Back Propagation of Gradients and Updating of Parameters*

We minimize $-L(\boldsymbol{\theta}_N)$ through gradient descent with the following steps: first, $-L(\boldsymbol{\theta}_N)$ is obtained from (25) based on MLE. Second, the gradients of $-L(\boldsymbol{\theta}_N)$ with respect to parameters of JDAN are formed as

$$\nabla_{[\boldsymbol{W}^+, \boldsymbol{B}]}[-L(\boldsymbol{\theta}_N)] = \nabla_{\boldsymbol{W}^+}[-L(\boldsymbol{\theta}_N)] + \nabla_{\boldsymbol{B}}[-L(\boldsymbol{\theta}_N)]. \quad (26)$$

Third, according to the chain rule of derivative, the gradients of $-L(\boldsymbol{\theta}_N)$ with respect to parameters of NFN are calculated as

$$\nabla_{\boldsymbol{\theta}_N}[-L(\boldsymbol{\theta}_N)] = \nabla_{\boldsymbol{\theta}_N}\boldsymbol{W}^+ \cdot \nabla_{\boldsymbol{W}^+}[-L(\boldsymbol{\theta}_N)] + \nabla_{\boldsymbol{\theta}_N}\boldsymbol{B} \cdot \nabla_{\boldsymbol{B}}[-L(\boldsymbol{\theta}_N)]. \quad (27)$$

Finally, Adam optimizer is chosen to update NFN with $\nabla_{\boldsymbol{\theta}_N}[-L(\boldsymbol{\theta}_N)]$, thus we pass the real gradients [with respect to $\boldsymbol{\theta}_N$] to NFN and update the parameters of NFN and JDAN.

The whole dataset is separated into training dataset, validation dataset, and testing dataset. If the mean value of the objective function on the training dataset has been greater than that on the validation dataset over 20 successive epochs, the training process will be stopped to avoid overfitting (set batch size as 32 and learning rate as 0.001).

## V. CASE STUDIES

### A. Setting of Numerical Simulation

The proposed model was validated on a modified New England 39-bus system as shown in Fig. 5 [4]. There are 3 regions in this system, and two wind farms with capacities of 400MW are connected at bus 17 and 21, respectively. Three power flowgates are determined by the tie-lines that connect different regions, i.e.: 1-39, 2-3, 18-3, and 16-15 (flowgate 1); 1-39, 2-3, 18-3, and 17-16 (flowgate 2); 15-16 and 17-16 (flowgate 3). $SM_{t+\tau}^1$, $SM_{t+\tau}^2$, and $SM_{t+\tau}^3$ are their corresponding security margins.

We used the probabilistic tool proposed in [4] to obtain the available information set ($X_t$) and the corresponding forecasting target ($SM_{t+\tau}^*$). Specifically, lead step $\tau$ was set as 15 min. The features in $X_t$ include bus voltages, power injections, and power flows, which are random time series from Monte-Carlo sampling with a time interval of 15 min. $SM_{t+\tau}^*$ was calculated based on state variables collected at time spot $t+\tau$ from the continuation power flow [7]. The number of samples in the whole dataset generated by the probabilistic tool is 15119, of which the first 40% was used for training, the middle 20% for validation, and the last 40% for testing. A grid-search method was adopted to find the optimal structure of JDAN-NFN (associated with $N_N$, $N_J$, $W_N$, and $W_J$) and the lag interval $\delta$. In detail, the optimal combination of $\{N_N\ N_J\ W_N\ W_J\ \delta\}$ is determined when the mean value of the objective function on the validation dataset is highest during the grid-search process, which is demonstrated in Appendix D.

### B. Prediction Results of the Proposed Method

The prediction results of JDAN-NFN are demonstrated in Fig. 6. Since we cannot show the high dimensional distribution directly, the forecasted JPDFs of multiple security margins over 96-time spots are transformed into conditional PDFs (CPDFs) of a single security margin. Specifically, the forecasted CPDF of one security margin is calculated as

$$\hat{f}_{C_i}(SM_{t+\tau}^i \mid X_t) = \frac{\Phi_J^{(N)}(SM_{t+\tau}; W^+, B)}{\Phi_{Ji}^{(N-1)}(SM_{t+\tau}; W^+, B)}, \quad (28)$$

where $\hat{f}_{C_i}(\cdot \mid X_t)$ is CPDF of $SM_{t+\tau}^i$ given any $SM_{t+\tau}^k \in SM_{t+\tau}$, $k \neq i$. $\Phi_{Ji}(SM_{t+\tau}; W^+, B)$ denotes $\lim_{SM_{t+\tau}^i \to +\infty} \Phi_J(SM_{t+\tau}; W^+, B)$.

The forecasted CPDFs $\hat{f}_{C_1}(\cdot \mid X_t)$, $\hat{f}_{C_2}(\cdot \mid X_t)$, and $\hat{f}_{C_3}(\cdot \mid X_t)$ over 96-time spots are demonstrated in Fig. 6(a), (b), and (c), respectively. The colormap on the right denotes the probability density, in which darker color means greater density while lighter color means smaller density. The green solid lines in each subfigure denote 96-time-spot observations of $SM_{t+\tau}^1$, $SM_{t+\tau}^2$, and $SM_{t+\tau}^3$, respectively. One can see that, in all the three subfigures, most observations lie on the areas with high probability densities. This illustrates that these forecasted CPDFs are very consistent with the reality. Therefore, we may reasonably conclude that the forecasted JPDF by JDAN-NFN has satisfactorily estimated the real JPDF of future security margins.

### C. Comparison of JDAN-NFN with Other Alternatives

Several multivariate density forecast models were implemented for comparison, including MKDE based on multivariate Gaussian kernel [16], multivariate Archimedean copulas of Clayton family (ACC) with parameter $\theta_C$ ($\theta_C = \{0.5, 1, 1.5\}$), and multivariate Archimedean copulas of

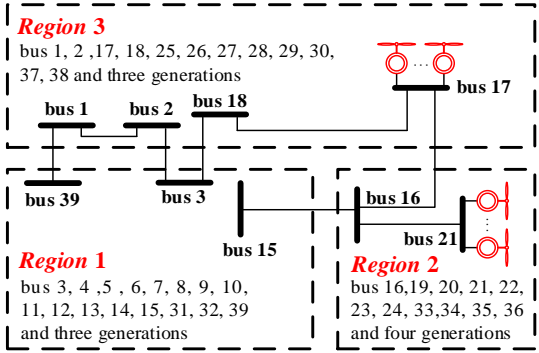

Fig. 5. The modified New England 39-bus system.

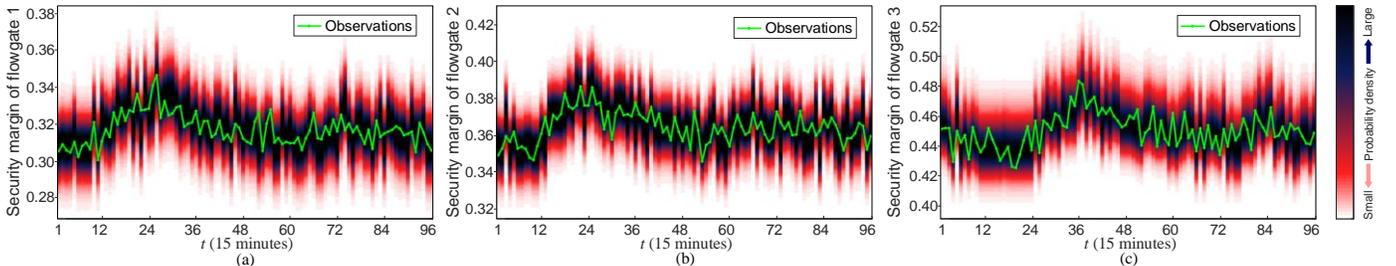

Fig. 6. Demonstration of the forecasted CPDFs.

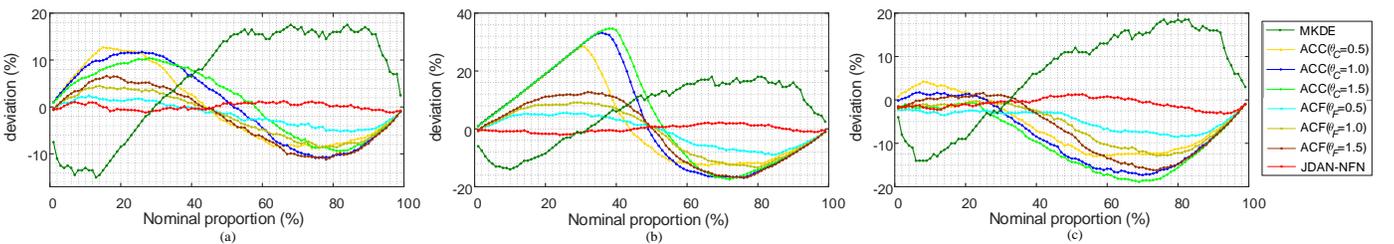

Fig. 7. Reliability evaluations. (a) reliability evaluation for CCDF$_1$. (b) reliability evaluation for CCDF$_2$. (c) reliability evaluation for CCDF$_3$.

Frank family (ACF) with parameter $\theta_F$ ( $\theta_F = \{0.5, 1, 1.5\}$ ) [17]. Copulas are JCDFs calculated from marginal CDFs, and marginal CDFs were obtained from DAN-NFN here [13]. The performance of all these approaches as well as JDAN-NFN were tested by the reliability evaluation [23], of which results are shown in Fig. 7.

Reliability is the foremost concern for probabilistic forecast model evaluation, which measures the deviations between the predicted distribution and the 'perfect reliability' case (the reality) at different quantiles. A treatment[1] here is that we transformed the forecasted joint distributions into conditional distributions [as we did in (28)] for the reliability evaluation. Denote the corresponding conditional CDF (CCDF) of $\hat{f}_{C_i}(\cdot | X_t)$ as $\hat{F}_{C_i}(\cdot | X_t)$ and $\hat{q}_{t+\tau|t}^{\alpha_j}$ as the quantile of $\hat{F}_{C_i}(\cdot | X_t)$ with nominal proportion $\alpha_j$ ( $j \in [1, J]$, $J$=99). From $\alpha_1$ =1% to $\alpha_J$ =99% with steps 1%, the deviation with different $\alpha_j$ can be obtained as

$$b_\tau^{\alpha_j} = \alpha_j - \frac{1}{N_\omega} \sum_{i=1}^{N_\omega} H(\hat{q}_{i+\tau|i}^{\alpha_j} - SM_{i+\tau}^*), \quad (29)$$

where $N_\omega$ is the number of samples in the testing dataset, $b_\tau^{\alpha_j}$ is the deviation corresponding to $\alpha_j$, and $H$ is the unit step function

$$H(x) = \begin{cases} 1, & x \geq 0 \\ 0, & x < 0 \end{cases}. \quad (30)$$

The reliability evaluations for different CCDFs are shown in each subfigure of Fig. 7. Additionally, an overview index $\bar{b}_\tau$ for reliability evaluation is calculated as $\bar{b}_\tau = (1/J)\sum_{j=1}^{J}\left|b_\tau^{\alpha_j}\right|$, which is presented in Table I. Illustration of these results are as follows.

TABLE I
OVERALL PERFORMANCE DEMONSTRATION

| Approach | | CCDF$_1$ | CCDF$_2$ | CCDF$_3$ |
|---|---|---|---|---|
| | | $\bar{b}_\tau$ % | $\bar{b}_\tau$ % | $\bar{b}_\tau$ % |
| MKDE | | 11.73 | 10.90 | 10.89 |
| ACC | $\theta_C = 0.5$ | 6.79 | 11.59 | 7.11 |
| | $\theta_C = 1.0$ | 7.31 | 14.70 | 8.44 |
| | $\theta_C = 1.5$ | 6.39 | 14.91 | 9.37 |
| ACF | $\theta_F = 0.5$ | 2.37 | 4.55 | 4.70 |
| | $\theta_F = 1.0$ | 4.20 | 7.29 | 5.63 |
| | $\theta_F = 1.5$ | 5.59 | 9.61 | 6.79 |
| JDAN-NFN | | **0.62** | **1.05** | **1.22** |

The reliability evaluation for CCDF $\hat{F}_{C_1}(\cdot | X_t)$ (denoted as CCDF$_1$) is analyzed first. As shown in Fig. 7(a), obvious deviations are observed in MKDE, ACCs, and almost all ACFs (except ACF with $\theta_F = 0.5$). MKDE tends to overestimate the quantiles with nominal proportions under 29% and overestimate the ones over 29%. ACCs and ACFs show different patterns from MKDE, they underestimate the quantiles with smaller nominal proportions (range from 1% to 58%) and overestimate the ones with bigger nominal proportions (range from 44% to 99%). ACF with $\theta_F = 0.5$ performs better than other copula-based methods, but non-negligible deviations still exist. JDAN-NFN has got the least $\bar{b}_\tau$ (0.62%), which demonstrates the highest reliability and the prominent superiority over other models.

---
[1]For univariate distribution with a nominal proportion, the quantile is a single value, which can be easily determined by the inverse function of CDF. While in the joint distribution, such quantile is a surface (i.e. the quantile surface). Joint distributions thus cannot be directly evaluated with reliability.

Similar analyses can be concluded in the reliability evaluations for CCDF$_2$ and CCDF$_3$ from the rest subfigures in Fig. 7 and Table I, we briefly introduce them here. As shown in Fig. 7(b) and (c), conspicuous over or underestimations can be checked in the reliability evaluations of MKDE and copula-based methods, and deviations with different nominal proportions in MKDE show different bias trends from that of copula-based ones. JDAN-NFN shows the lowest absolute deviation on the whole and achieves the least $\bar{b}_\tau$ (1.05% and 1.22%) in the reliability evaluations for both CCDF$_2$ and CCDF$_3$.

Summarizing the above analyses, one can see that JDAN-NFN has exhibited very competitive and much preferable performance compared with the existing multivariate density forecast models.

### D. Comparison of $\Omega$ with the Security Margin

As shown in Table II, the deterministic security assessment index $\Omega$ was compared with the observation of ($SM_{t+\tau}^1$, $SM_{t+\tau}^2$, $SM_{t+\tau}^3$) under different operation conditions (different samples). The security thresholds $\gamma_1$, $\gamma_2$, and $\gamma_3$ were set as 0.7, 0.65, and 0.6, respectively. Thus, when assessing the system security by security margins, the operation condition is secure if $SM_{t+\tau}^1 \geq$ 0.3, $SM_{t+\tau}^2 \geq 0.35$, and $SM_{t+\tau}^3 \geq 0.4$ in the observation (also named as a secure observation). We randomly chose several operation conditions in the test dataset. For each chosen operation condition, 1000 scenarios of ($SM_{t+\tau}^1$, $SM_{t+\tau}^2$, $SM_{t+\tau}^3$) were generated by Monte Carlo sampling from the predicted distribution of security margins. Whether the generated scenario is secure or not was determined by security thresholds as well. The analyses about $\Omega$, the observation of security margins, and the proportion of secure scenarios[2] under different operation conditions in Table II are as follows.

TABLE II
COMPARISON OF $\Omega$ UNDER DIFFERENT OPERATION CONDITIONS

| Operation condition | Observation of ($SM_{t+\tau}^1$, $SM_{t+\tau}^2$, $SM_{t+\tau}^3$) | Proportion of secure scenarios (%) | $\Omega$ (%) |
|---|---|---|---|
| 1 | (0.338, 0.376, 0.467) | 95.5 | 94.9 |
| 2 | (0.342, 0.386, 0.482) | 60.7 | 60.8 |
| 3 | (0.301, 0.345, 0.425) | 93.8 | 93.1 |
| 4 | (0.306, 0.349, 0.431) | 41.1 | 41.8 |
| 1$^+$ | (0.317, 0.360, 0.445) | 46.0 | 45.4 |
| 2$^+$ | (0.293, 0.333, 0.405) | 92.2 | 90.1 |
| 3$^+$ | (0.346, 0.390, 0.470) | 95.8 | 94.3 |
| 4$^+$ | (0.310, 0.356, 0.437) | 82.9 | 81.0 |

The wind power outputs of operation condition with no superscript are shown by Fig. 8(a) in Appendix E. The operation conditions labeled with superscript "+" mean that the wind was strong, which yielded more volatile wind power outputs [see Fig. 8(b) in Appendix E], and JDAN-NFN was retained by the newly generated data in these cases.

---
[2]The proportion of secure scenarios in generated scenarios demonstrates the possibility of the secure operation condition during a future period.

We see that observations of security margins show no strong correlation with the proportion of secure scenarios, such observations are thus information-limited for power system operations with uncertainties. Specifically, significantly different scenarios can be generated with similar observations (such as operation conditions 3 and 4), which may be explained as that the predicted distributions of security margins are very different under these conditions. Paradoxes can also be observed between the observation of security margins and generated scenarios: the secure observations correspond low proportion of secure scenarios (operation conditions 2 and $1^+$), or insecure observation corresponds high proportion of secure scenarios (operation condition $2^+$). $\Omega$ accords well with the proportion of secure scenarios in all cases, which reveals the distribution information of future security margins under different operation conditions. It is thus more informative for the power system security assessment. Moreover, the average computation time of $\Omega$ for each time spot is 0.0492s, which is suitable for online applications.

*E. Discussion for the Practical Application*

In practical applications, datasets are generated by the probabilistic tool according to the day-ahead dispatch schedule first. Then, JDAN-NFN will be trained and prepared for online security assessment the next day. Therefore, the procedure of datasets generation and model training can both be boosted by high-performance computing clusters in off-line mode. Furthermore, operators may just care about several flowgates locally, so sub-models can be trained in parallel involving different sub-combinations of flowgates in the system, which degenerates the complexity of JDAN-NFN model, and accelerates the training process.

## VI. CONCLUSION

In this paper, a novel multivariate density forecast model JDAN-NFN has been developed for online power system security assessment with uncertainties. In JDAN-NFN, the input-output function of JDAN approximates the real JCDFs of security margins, and NFN forecasts the weights and biases of JDAN. NFN and JDAN both possess deep residual architectures for large capacities, and the proposed model is trained based on MLE. Then, a deterministic security assessment index $\Omega$ is proposed to show the future security of power system operations. Numerical tests of a typical power system have demonstrated the superiority of JDAN-NFN and illustrated that $\Omega$ accords well with the distribution information of future security margins.

## APPENDIX

*A. Explanation for Why a Normal MISO Positive-weighted ANN Cannot be Used to Represent the JCDF*

For conciseness, denote $\mathbf{y} = [y_1, y_2, \cdots, y_n]$ as the input, which represents a vector with $n$ variables. We construct the mapping function of a positive-weighted ANN as

$$\Phi(\mathbf{y}; \mathbf{w}^+, \mathbf{b}) = z\{\mathbf{w}_{K+1}^+ \cdots z[\mathbf{w}_2^+ \cdot z(\mathbf{w}_1^+ \cdot \mathbf{y} + \mathbf{b}_1) + \mathbf{b}_2] \cdots + \mathbf{b}_{K+1}\}, \quad (31)$$

where $K$ is the number of hidden layers. $z$ is the activation function (sigmoid, tanh, linear, or ReLU), and can be different in different layers. Tensors $\mathbf{w}^+$ and $\mathbf{b}$ represent all the weights and biases, respectively, which are determined by the outputs of NFN. Tensors $\mathbf{w}_k^+ \in \mathbf{w}^+$ and $\mathbf{b}_k \in \mathbf{b}$, $k \in [1, K+1]$, are weights and biases in $k$th layer, respectively.

For $k=1, \cdots, K$, define the input-output mapping function of layer $k$ in (31) as

$$\mathbf{Y}_k = z(\mathbf{w}_k^+ \cdot \mathbf{Y}_{k-1} + \mathbf{b}_k) = \begin{bmatrix} z(\mathbf{w}_{k,1}^+ \cdot \mathbf{Y}_{k-1} + \mathbf{b}_{k,1}) \\ \vdots \\ z(\mathbf{w}_{k,l}^+ \cdot \mathbf{Y}_{k-1} + \mathbf{b}_{k,l}) \\ \vdots \end{bmatrix} = \begin{bmatrix} z_{k,1} \\ \vdots \\ z_{k,l} \\ \vdots \end{bmatrix}, \quad (32)$$

$$\mathbf{Y}_{K+1} = z(\mathbf{w}_{K+1}^+ \cdot \mathbf{Y}_K + \mathbf{b}_{K+1}), \quad (33)$$

where $\mathbf{Y}_0 = \mathbf{y}$. Tensors $\mathbf{w}_{k,l}^+$ and $\mathbf{b}_{k,l}$ represent the $l$th row in $\mathbf{w}_k^+$ and $\mathbf{b}_k$, respectively. $z_{k,l}$ is short for $z(\mathbf{w}_{k,l}^+ \cdot \mathbf{Y}_{k-1} + \mathbf{b}_{k,l})$. Note that $\mathbf{w}_{K+1}^+$ and $\mathbf{b}_{K+1}$ in the output layer are vector and scalar, respectively. Then, for $k=1, \cdots, K$:

$$\frac{d\mathbf{Y}_k}{d\mathbf{Y}_{k-1}} = \begin{bmatrix} z'_{k,1} \cdot (\mathbf{w}_{k,1}^+)^T & \cdots & z'_{k,l} \cdot (\mathbf{w}_{k,l}^+)^T & \cdots \end{bmatrix}, \quad (34)$$

$$\frac{d\mathbf{Y}_{K+1}}{d\mathbf{Y}_K} = z'_{K+1}(\mathbf{w}_{K+1}^+ \cdot \mathbf{Y}_K + \mathbf{b}_{K+1}) \cdot (\mathbf{w}_{K+1}^+)^T, \quad (35)$$

Thus,

$$\frac{d\Phi(\mathbf{y}; \mathbf{w}^+, \mathbf{b})}{d\mathbf{y}} = \prod_{k=1}^{K+1} \frac{d\mathbf{Y}_k}{d\mathbf{Y}_{k-1}}. \quad (36)$$

The first derivatives of different activation function $z$ are

$$z' = \begin{cases} \sigma \cdot (1-\sigma), & \text{if } z \text{ is sigmoid} \\ 2\sigma \cdot (1-\sigma), & \text{if } z \text{ is tanh} \\ 1, & \text{if } z \text{ is linear} \\ 0 \text{ or } 1, & \text{if } z \text{ is ReLU} \end{cases}. \quad (37)$$

Since $\sigma \in (0,1)$, $z'$ is nonnegative for all kinds of activation functions considered in this paper. Considering that every entry in $\mathbf{w}^+$ is positive and combining (34)-(37), one can verify that $d\Phi(\mathbf{y}; \mathbf{w}^+, \mathbf{b})/d\mathbf{y}$ is a $n$ dimensional vector, and any entry in it is positive. Therefore, $\Phi(\mathbf{y}; \mathbf{w}^+, \mathbf{b})$ is multivariate monotone non-decreasing.

Now, we illustrate that why such multivariate monotone non-decreasing property cannot be extended to the higher-order-derivative form that meets condition (ii). For conciseness, an example is taken when there are two layers in the positive-weighted ANN, which can be denoted as

$$\Gamma = z[\mathbf{w}_2^+ \cdot z(\mathbf{w}_1^+ \cdot \mathbf{y} + \mathbf{b}_1) + \mathbf{b}_2]. \quad (38)$$

Define $y_p$ and $y_q$ are arbitrary two entries in $\mathbf{y}$, based on the analyses about (31)-(36), the second partial derivative of $\Gamma$ with respect to them can be derived as

$$\frac{\partial^2 \Gamma}{\partial y_p \partial y_q} = \begin{bmatrix} z''_{1,1} \cdot \mathbf{w}_{1,1p}^+ \cdot \mathbf{w}_{1,1q}^+ & \cdots & z''_{1,l} \cdot \mathbf{w}_{1,lp}^+ \cdot \mathbf{w}_{1,lq}^+ & \cdots \end{bmatrix}$$
$$\cdot z'(\mathbf{w}_2^+ \cdot \mathbf{Y}_2 + \mathbf{b}_2) \cdot (\mathbf{w}_2^+)^T$$
$$+ \left\{ \begin{bmatrix} z'_{1,1} \cdot \mathbf{w}_{1,1p}^+ & \cdots & z'_{1,l} \cdot \mathbf{w}_{1,lp}^+ & \cdots \end{bmatrix} \cdot (\mathbf{w}_2^+)^T \right\}$$
$$\cdot \left\{ \begin{bmatrix} z'_{1,1} \cdot \mathbf{w}_{1,1q}^+ & \cdots & z'_{1,l} \cdot \mathbf{w}_{1,lq}^+ & \cdots \end{bmatrix} \cdot z''(\mathbf{w}_2^+ \cdot \mathbf{Y}_2 + \mathbf{b}_2) \cdot (\mathbf{w}_2^+)^T \right\}$$
$$, (39)$$

where $\mathbf{w}_{1,lp}^+$ ($\mathbf{w}_{1,lq}^+$) is the element at $l$th row and $p$th ($q$th) column in $\mathbf{w}_1^+$. The second derivatives of different activation function $z$ are

$$z'' = \begin{cases} \sigma \cdot (1-\sigma) \cdot (1-2\sigma) & \text{if } z \text{ is sigmoid} \\ 2\sigma \cdot (1-\sigma) \cdot (1-2\sigma) & \text{if } z \text{ is tanh} \\ 0 & \text{if } z \text{ is linear or ReLU} \end{cases} \quad (40)$$

It shows that the nonnegative property does not always hold for $z''$. For sigmoid or tanh activation function, $z''$ will be negative if the intermittent computing result is greater than zero when doing forward or backward propagation in the network, which is observed very commonly. Although $z''$ can be nonnegative all the time for ReLU or linear activation function, the NN still could learn nothing because the gradients are always zero. Combine (39) and (40), condition (ii) cannot be guaranteed.

A simple idea is to find a very special activation function so that

$$z' \geq 0, z'' \geq 0, \cdots, z^{(n)} \geq 0, \quad (41)$$

which ensures condition (ii). One activation function satisfying (41) is exponential function ($e^x$). However, it is rarely used in NNs as exponential function may suffer from vanishing/exploding gradient problems.

Based on the analyses above, a normal MISO positive-weighted ANN cannot be used to represent the JCDF.

### B. Example for Variable Independency

We discard the coupling layer, then construct the forecasted joint distribution $\hat{\phi}'_{t+\tau}|_{X_t}$ with JCDF $\hat{F}'(\cdot|X_t)$ as

$$\hat{F}'(SM_{t+\tau}|X_t) = \Phi'_J(SM_{t+\tau}; W^+, B) = \prod_{i=1}^{N} \bar{\Phi}^i. \quad (42)$$

Set $A = \{i | i \in [1, N]\}$, the marginal PDF of $SM^i_{t+\tau}$ can be calculated as

$$\hat{f}'_{M_i}(SM^i_{t+\tau}|X_t) = \frac{\partial \left( \lim_{\substack{SM^j_{t+\tau} \to +\infty \\ j \in A \setminus \{i\}}} \Phi'_J(SM_{t+\tau}; W^+, B) \right)}{\partial SM^i_{t+\tau}} = \frac{\partial \bar{\Phi}^i}{\partial SM^i_{t+\tau}}. \quad (43)$$

So, the JPDF is formulated as

$$\hat{f}'(SM_{t+\tau}|X_t) = \prod_{i=1}^{N} \hat{f}'_{M_i}(SM^i_{t+\tau}|X_t), \quad (44)$$

which implies that the joint distribution is formed by the multiplication of marginal distributions, so variables are independent in the joint distribution.

### C. JDAN-NFN, an Ideal Multivariate Density Forecast Model

The proof of ideal multivariate density forecast model for JDAN-NFN is similar to that of the univariate one in [13]. According to the definition of *ideal density forecast model* in (1), we need to prove that $\forall X_t \in E, D_{X_t} = \Xi$. To obtain the $D_{X_t}$ of JDAN-NFN, we should first determine the output of NFN. NFN can be presented as

$$[W^+, B] = \Phi_N(X_t; \theta_N), \quad (45)$$

where $\theta_N$ represents parameters of NFN. $\Phi_N(\cdot; \theta_N)$ is the input-output mapping function of NFN. As the capacity of NFN can be very large with deep structure, one can thus assume that for any given $X_t$, $[W^+, B]$ is only constrained by the range of their activation functions. In other words, for any $W^+ \in \mathbb{R}^+$ and $B \in \mathbb{R}$, one can always find $[W^+, B] = \Phi_N(X_t; \theta_N)$ regardless of $X_t$. Therefore, given $X_t$, the range of NFN (denoted as $D_N$) is

$$D_N = \{[W^+, B] | W^+ \in \mathbb{R}^+, B \in \mathbb{R}\}. \quad (46)$$

Combining (46) and (13), $D_{X_t}$ can be presented as

$$D_{X_t} = \{\hat{\phi}_{t+\tau}|_{X_t} \text{ with JCDF } \hat{F}(\cdot|X_t) | [W^+, B] \in D_N\}. \quad (47)$$

To prove $D_{X_t} = \Xi$, two steps are as follows.

*Step 1*: Prove that $\forall X_t \in E, \Xi \subseteq D_{X_t}$.

Define the joint distribution $\tilde{\phi}_{t+\tau}|_{X_t} \in \Xi$ with JCDF as $\tilde{F}(\cdot|X_t)$. $\tilde{F}(\cdot|X_t)$ is continuous and multivariate monotone non-decreasing, and

$$\lim_{SM_{t+\tau} \to +\infty} \tilde{F}(SM_{t+\tau}|X_t) = 1, \quad (48)$$

$$\lim_{SM^i_{t+\tau} \to -\infty} \tilde{F}(SM_{t+\tau}|X_t) = 0, \quad \forall i \in [1, N]. \quad (49)$$

To prove $\tilde{\phi}_{t+\tau}|_{X_t} \in D_{X_t}$, one only need to prove that there exists a estimation $\hat{F}(SM_{t+\tau}|X_t)$ with $[\hat{W}^+, \hat{B}] \in D_{X_t}$ that is equal to $\tilde{F}(SM_{t+\tau}|X_t)$ for all $SM_{t+\tau} \in (-\infty, +\infty)$. As JDAN is a positive-weighted ANN, according to *Theorem 1*, JDAN can approximate any multivariate monotone function by enlarging its capacity. Thus, we can always find a combination $[\hat{W}^+ \in \mathbb{R}^+, \hat{B} \in \mathbb{R}]$ that meets

$$\Phi_J(SM_{t+\tau}; \hat{W}^+, \hat{B}) = \tilde{F}(SM_{t+\tau}|X_t), \forall SM_{t+\tau} \in (-\infty, +\infty). \quad (50)$$

As the activation function of JDAN is sigmoid and linear, $\Phi_J(\cdot; \hat{W}^+, \hat{B})$ is surely continuous. So we can conclude that

$$\lim_{SM_{t+\tau} \to +\infty} \Phi_J(SM_{t+\tau}; \hat{W}^+, \hat{B}) = \lim_{SM_{t+\tau} \to +\infty} \tilde{F}(SM_{t+\tau}|X_t) = 1, \quad (51)$$

$$\lim_{SM^i_{t+\tau} \to -\infty} \Phi_J(SM_{t+\tau}; \hat{W}^+, \hat{B}) = \lim_{SM^i_{t+\tau} \to -\infty} \tilde{F}(SM_{t+\tau}|X_t) = 0, \forall i \in [1, N]. \quad (52)$$

Substitute $W^+$ and $B$ in (13) by $\hat{W}^+$ and $\hat{B}$, respectively, then from (48)-(52) we have

$$\hat{F}(SM_{t+\tau}|X_t) = \tilde{F}(SM_{t+\tau}|X_t), \forall SM_{t+\tau} \in (-\infty, +\infty). \quad (53)$$

$$\lim_{SM_{t+\tau} \to +\infty} \hat{F}(SM_{t+\tau}|X_t) = \lim_{SM_{t+\tau} \to +\infty} \tilde{F}(SM_{t+\tau}|X_t) = 1. \quad (54)$$

$$\lim_{SM^i_{t+\tau} \to -\infty} \hat{F}(SM_{t+\tau}|X_t) = \lim_{SM^i_{t+\tau} \to -\infty} \tilde{F}(SM_{t+\tau}|X_t) = 0. \quad (55)$$

Therefore, for any $SM_{t+\tau}$, there exists $\hat{F}(SM_{t+\tau}|X_t)$ with $[\hat{W}^+, \hat{B}] \in D_{X_t}$ that is equal to $\tilde{F}(SM_{t+\tau}|X_t)$, thus $\tilde{\phi}_{t+\tau}|_{X_t} \in D_{X_t}$. As $\tilde{\phi}_{t+\tau}|_{X_t}$ can be any element in $\Xi$, so $\Xi \subseteq D_{X_t}$.

*Step 2*: Prove that $\forall X_t \in E, D_{X_t} \subseteq \Xi$.

For any estimated joint distribution $\hat{\phi}_{t+\tau}|_{X_t} \in D_{X_t}$ with JCDF, it can be denoted as $\hat{F}(SM_{t+\tau}|X_t)$ parameterized by $[\hat{W}^+, \hat{B}] \in D_N$. Substitute $W^+$ and $B$ in (13) by $\hat{W}^+$ and $\hat{B}$, respectively, then we have

$$\hat{F}(SM_{t+\tau}|X_t) = \Phi_J(SM_{t+\tau}; \hat{W}^+, \hat{B}). \quad (56)$$

Substitute $[\boldsymbol{W}^+, \boldsymbol{B}]$ in (10)-(12) by $[\hat{\boldsymbol{W}}^+, \hat{\boldsymbol{B}}]$, then we can infer that $\Phi_J(\cdot; \hat{\boldsymbol{W}}^+, \hat{\boldsymbol{B}})$ is continuous and

$$\Phi_J^{(N)}(\boldsymbol{SM}_{t+\tau}; \hat{\boldsymbol{W}}^+, \hat{\boldsymbol{B}}) \geq 0, \quad (57)$$

$$\lim_{\boldsymbol{SM}_{t+\tau} \to +\infty} \Phi_J(\boldsymbol{SM}_{t+\tau}; \hat{\boldsymbol{W}}^+, \hat{\boldsymbol{B}}) = 1, \quad (58)$$

$$\lim_{\boldsymbol{SM}^i_{t+\tau} \to -\infty} \Phi_J(\boldsymbol{SM}_{t+\tau}; \hat{\boldsymbol{W}}^+, \hat{\boldsymbol{B}}) = 0, \quad \forall i \in [1, N]. \quad (59)$$

Thus, $\hat{F}(\cdot | \boldsymbol{X}_t)$ satisfies (5), then $\hat{\boldsymbol{\phi}}_{t+\tau}|_{\boldsymbol{X}_t} \in \Xi$. As $\hat{\boldsymbol{\phi}}_{t+\tau}|_{\boldsymbol{X}_t}$ can be any element in $\boldsymbol{D}_{\boldsymbol{X}_t}$, so $\boldsymbol{D}_{\boldsymbol{X}_t} \subseteq \Xi$.

Combining step 1 and 2, we can finally prove that $\forall \boldsymbol{X}_t \in \boldsymbol{E}, \boldsymbol{D}_{\boldsymbol{X}_t} = \Xi$. Therefore, JDAN-NFN is an ideal multivariate density forecast model if the capacities of JDAN and NFN are built large enough.

*D. Grid-search Process*

The training of JDAN-NFN was implemented on CentOS 7.6 with 8 TITAN V GPUs, and exhaustive research was carried out to determine the optimal combination $\{N_N\ N_J\ W_N\ W_J\ \delta\}$, in which $N_N$ was chosen from $\{2\ 4\ 8\ 16\}$, $N_J$ from $\{2\ 4\ 8\ 16\}$, $W_N$ from $\{16\ 32\ 64\ 128\}$, $W_J$ from $\{16\ 32\ 64\ 128\}$, $\delta$ from $\{5\ \text{min}\ 10\ \text{min}\ 15\ \text{min}\ 20\ \text{min}\ 25\ \text{min}\}$. The optimal $\{N_N\ N_J\ W_N\ W_J\ \delta\}$ are listed in Table III.

TABLE III
OPTIMAL COMBINATION OF HYPER-PARAMETERS

| $N_N$ | $N_J$ | $W_N$ | $W_J$ | $\delta$ /min |
|---|---|---|---|---|
| 8 | 4 | 64 | 64 | 20 |

*E. Wind Farm Outputs at Bus 17 and 21*

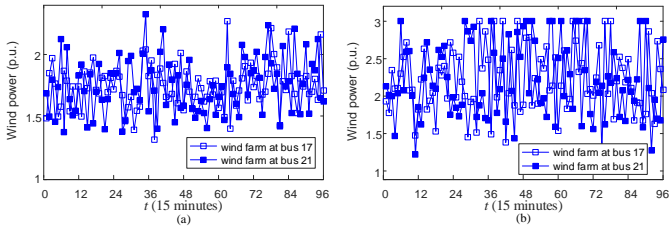

Fig. 8. The wind farm outputs (aggregation of wind generators).